
\input harvmac
\noblackbox
\def \eq#1 {\eqno {(#1)}}
\def \cB {{\cal B}}
\def \cA {{\cal A}}
\def \four{{\textstyle{1\over 4}}}


\def\a{\alpha}

\def\g{\gamma}

\def\d{\delta}

\def\e{\epsilon}

\def\p{\phi}

\def\m{\mu}
\def\n{\nu}

\def\l{\lambda}

\def\s{\sigma}
\def\cA{{\cal A}}
\def \sm {$\s$-model\ }
\
\def \bd {\bar \del}

\def \ha {{1\over 2}}

\def \ov {\over}

\def \p {\phi}
\def \vp {\varphi}

\def \bd {\bar \del}

\def \a {\alpha}

\def \log {{\rm log \ }}
\def \ln {{\rm \ ln \  }}
\def \det {{\ \rm det \ }}

\def \l {\lambda}
\def \p {\phi}

\def \m {\mu }
\def \n {\nu}
\def \ep {\epsilon}
\def\g {\gamma}

\def \d {\delta}

\def \s {\sigma}
\def \t {\tau}

\def \e#1 {{{\rm e}^{#1}}}

\def \eq#1 {\eqno {(#1)}}
\def \sm {$\s$-model\ }

\def \bd  {{ \bar \del }}

\def \bd  { \bar \del }


\def \p {\phi}
\def \ep {\epsilon}
\def \s {\sigma}

\def \d {\delta}
\def \l {\lambda}
\def \m {\mu}

\def \g {\gamma}
\def \n {\nu}

\def \e#1 {{{\rm e}^{#1}}}
\def \vp {\varphi}

\def \m {\mu}  

\def \ep {\epsilon}

\def \eq#1 {\eqno{(#1)}}
\def \e {\rm e}

\def \e#1 {{\rm e}^{#1}}

\def \ln { {\rm ln } }

\def \l {\lambda}
\def \p {\phi}
\def \vp {\varphi}
\def  \g {\gamma}

\def\({\left (}
\def\){\right )}
\def\[{\left [}
\def\]{\right ]}

\def\bd {{\bar \del}}
\def \vp {\varphi}

\def \eq#1 {\eqno{(#1)}}

\def \a {\alpha}

\def \p {\phi}
\def \ep {\epsilon}
\def \s {\sigma}

\def \d {\delta}
\def \l {\lambda}
\def \m {\mu}
\def \g {\gamma}
\def \n {\nu}

\def \e#1 {{\rm e}^{#1}}

\def \vp {\varphi}

\def \ha { { 1\over 2 }}

\def \ov {\over}

\def \sm  { sigma model\ }

\def\np {  Nucl. Phys. }
\def \pl { Phys. Lett. }
\def \mpl { Mod. Phys. Lett. }
\def \prl { Phys. Rev. Lett. }
\def \pr  { Phys. Rev. }

\def \ijmp { Int. J. Mod. Phys. }
\def \cqg {Class. Quant. Grav.}

\lref \hhtt{ J. Horne, G. Horowitz and A. Steif, \jnl \prl,  68, 568, 1992.}

 \lref \busc { T.H. Buscher, \pl B194(1987)59 ; \pl B201(1988)466.}
\lref \pan  { J. Panvel, \pl B284(1992)50. }

\lref \mye {  R. Myers, \pl B199(1987)371;
    I. Antoniadis, C. Bachas, J. Ellis, D. Nanopoulos,
\pl B211(1988)393;
 \np B328(1989)115. }

\lref  \horne { J.H. Horne and G.T. Horowitz, \np B368(1992)444.}

\lref \koki { K. Kounnas and  E. Kiritsis, preprint CERN-TH.7059/93;
hep-th/9310202. }

\lref \tsdu { A.A. Tseytlin, \pl B242(1990)163; \np B350(1991)395.  }

\def \lr { \lref}

\gdef \jnl#1, #2, #3, 1#4#5#6{ { #1~}{ #2} (1#4#5#6) #3}

\lr \ghrw{J. Gauntlett, J. Harvey, M. Robinson, and D. Waldram,
\jnl \np, B411, 461, 1994.}
\lr \garf{D. Garfinkle, \jnl \pr, D46, 4286, 1992.}

\lref \tspl {A.A. Tseytlin, \jnl \pl, B317, 559, 1993.}
\lref \tssfet { K. Sfetsos and A.  Tseytlin, \jnl  \pr, D49, 2933, 1994.}
\lref \klts {C. Klim\v c\'\i k  and A. Tseytlin, ``Exact four dimensional
string solutions and Toda-like sigma models from null-gauged
WZNW models",  preprint
 Imperial/TP/93-94/17, hep-th/9402120.}

\lr \sfexac {K. Sfetsos,  \jnl \np, B389, 424,  1993.}

\lr \tsmac{A.A. Tseytlin, \jnl \pl,  B251, 530, 1990.}

\lr \cakh{C. Callan and R. Khuri, \jnl \pl, B261, 363, 1991;
R. Khuri, \jnl \np, B403, 335, 1993.}
\lr \dgt{M.J. Duff, G.W. Gibbons and P.K. Townsend, ``Macroscopic superstrings
as interpolating solitons", DAMTP/R-93/5, hep-th/9405124.}

\lref \ger {A. Gerasimov, A. Morozov, M. Olshanetsky, A. Marshakov and S.
Shatashvili, \jnl \ijmp,
A5, 2495,  1990. }

\lr \frts {E.S. Fradkin  and A.A. Tseytlin, \jnl \pl, B158, 316, 1985;
\jnl \np, B261, 1, 1985.}
\lr \tsred  {A.A. Tseytlin, \jnl  \pl, B176, 92, 1986; \jnl  \np, B276, 391,
 1986.}
\lr \grwi   { D. Gross and E. Witten, \jnl \np, B277, 1, 1986.}

\lr \gps {S.  Giddings, J. Polchinski and A. Strominger, \jnl  \pr,  D48,
 5784, 1993. }

\lr \tsmpl {A.A. Tseytlin, \jnl  \mpl, A6, 1721, 1991.}
\lr \vene { }
\lr \kltspl { C. Klim\v c\'\i k and A. Tseytlin, \jnl \pl, B323, 305, 1994.}
\lr \shwts { A. Schwarz and A. Tseytlin, \jnl \np, B399, 691, 1993.}
\lr \callnts { C. Callan and Z. Gan, \jnl  \np, B272, 647, 1986;  A.A.
Tseytlin,
\jnl \pl,
B178, 34, 1986.}

\lr \desa{ H. de Vega and N. Sanchez, \jnl
\pr, D45, 2783, 1992; \jnl \cqg, 10, 2007, 1993.}
\lr \desas{ H. de Vega and N. Sanchez, \jnl
\pl, B244,  215, 1990.}
\lref \tsnul { A.A. Tseytlin, \jnl \np, B390, 153, 1993.}

\lref \dunu { G. Horowitz and A. Steif, \pl B250 (1990) 49;
 E. Smith and J. Polchinski, \pl B263 (1991) 59. }

\lr \gauged {I. Bars and K. Sfetsos, \jnl  \mpl, A7, 1091, 1992;
 P. Ginsparg and F. Quevedo, \jnl \np, B385, 527, 1992. }
\lr \bsfet {I. Bars and K. Sfetsos, \jnl \pr, D46, 4510, 1992; \jnl \pr,
 D48, 844, 1993. }
\lr \tsnp{ A.A. Tseytlin, \jnl \np, B399, 601, 1993;  \jnl \np, B411, 509,
1994.}
\lr \gibb{A. Dabholkar, G.W. Gibbons, J. Harvey and F. Ruiz Ruiz, \jnl \np,
B340,
33, 1990.}
\lr \hhs{
G. Horowitz, in:  {\it  String Theory and Quantum Gravity '92,  Proc. of the
1992 Trieste Spring School}, ed.  J. Harvey et al. (World Scientific,
Singapore, 1993); hep-th/9210119.}

\lr \horstr{G. Horowitz and A. Strominger, \jnl \np, B360, 197, 1991.}

\lr \givkir {A.  Giveon and E. Kiritsis, \jnl \np, B411, 487, 1994.  }

\lr \jac{I. Jack and D. Jones, \jnl \pl, B200, 453, 1988.}
\lr \metts{R. Metsaev and A.A. Tseytlin, \jnl \np,  B293, 385, 1987.  }
\lr \horv{ P. Horava, \jnl \pl,
B278, 101, 1992.}

\lref \FT {E.S. Fradkin and A.A. Tseytlin, Phys.Lett. B158(1985)316; Nucl.Phys.
B261(1985)1. }
\lref \love {C. Lovelace, \pl B135(1984)75; \np B273(1986)413.}
\lref \sch {J. Scherk and J.H. Schwarz, \np B81(1974)118.}

\lref \por{ A. Giveon, M. Porrati and E. Rabinovici, preprint RI-1-94,
hep-th/9401139. }
\lref \klits{  C. Klim\v c\'\i k  and A.A. Tseytlin, unpublished (1994). }

\lr \tsmac{A.A. Tseytlin, \jnl \pl,  B251, 530, 1990.}

\lr \tspa { A.A. Tseytlin, ``Exact  string solutions
and duality", to appear in:  {\it Proceedings of the 2nd Journ\'ee Cosmologie,
 Observatoire de Paris, June 2-4, 1994}, ed. H. De Vega and N. Sanchez (World
Scientific, Singapore);  hep-th/9407099.}

\lr\bergsh { E. Bergshoeff, R. Kallosh and T. Ort\'in, \jnl \pr,  D47, 5444,
1993.}

\lr\sen{A. Sen, \jnl \np, B388, 457, 1992. }
\lr \garf{D. Garfinkle, \jnl \pr, D46, 4286, 1992.}
\lr \wald {D. Waldram, \jnl \pr, D47, 2528, 1993.}

\lr \born { M. Born, { Proc. Roy. Soc. } { A143} (1934) 410;
 M. Born and L. Infeld, { Proc. Roy. Soc. } { A144} (1934) 425.}

\lr \hrt  { G. Horowitz and A. Tseytlin, ``Extremal black holes as exact string
solutions", preprint  Imperial/TP/93-94/51, UCSBTH-94-24, hep-th/9408040.}

\lr \gib { G.W. Gibbons, \jnl \np, B207, 337, 1982. }

\lr \gim { G.W. Gibbons and K. Maeda,  \jnl \np, B298, 741, 1988. }
\lr\gar { D. Garfinkle, G. Horowitz and A. Strominger, \jnl \pr,  D43, 3140,
1991; {\bf D45} (1992) 3888(E). }
 \lr \witt{ E. Witten, \jnl \pr, D44, 314, 1991. }

\lr \guv {
 D. Amati and C. Klim\v c\'\i k,
\jnl \pl, B219, 443, 1989; G. Horowitz and A. Steif,  \jnl \prl, 64, 260,
1990.}

\lr \gurs{M. G\"urses, \jnl \pr, D46, 2522, 1992.}

\lr\bergsh { E. Bergshoeff, R. Kallosh and T. Ort\'in, \jnl \pr,  D47, 5444,
1993; E. Bergshoeff, I. Entrop and R. Kallosh,
\jnl \pr, D49, 6663, 1994.}
\lr \wald {D. Waldram, \jnl \pr, D47, 2528, 1993.}

\lr \callan { C. Callan, R. Myers and M. Perry, \jnl \np, B311, 673, 1988;
R. Myers, \jnl \np, B289, 701, 1987.}

\lr\rot{J.~Horne and G.~Horowitz, \jnl \pr, D46, 1340, 1992;
A. Sen, \jnl \prl, 69, 1006, 1992.}
\lr \ghrw{J. Gauntlett, J. Harvey, M. Robinson and D. Waldram,
\jnl \np, B411, 461, 1994.}
\lr \garf{D. Garfinkle, \jnl \pr, D46, 4286, 1992.}

\lr\tser { A. Tseytlin, ``String cosmology and dilaton",  in: {\it
String Quantum Gravity  and Physics at the Planck scale,
Proceedings of the 1992 Erice
workshop}, ed. N. Sanchez (World
Scientific, Singapore, 1993) p.202; hep-th/9206067; \jnl \ijmp, D1, 223, 1992,
hep-th/9203033. }
\lr \tscv{M. Cveti\v c  and A. Tseytlin, \jnl \np, B416, 137, 1994.}
\lr \poldam{ T. Damour and A. Polyakov, \jnl \np, B423, 532, 1994. }
\lr \harms{B. Harms and Y. Leblanc,
``Conjectures on nonlocal effects in string black holes",
hep-th/9307042.}
\lr \frtse{E.S. Fradkin and A.A. Tseytlin, \jnl \pl,  B163, 123, 1985. }

\lr \cakh{C. Callan and R. Khuri, \jnl \pl, B261, 363, 1991;
R. Khuri, \jnl \np, B403, 335, 1993.}
\lr \ghrw{J. Gauntlett, J. Harvey, M. Robinson and D. Waldram,
\jnl \np, B411, 461, 1994.}

\lr \chs {C. Callan, J. Harvey and A. Strominger, \jnl \np,  B359,
 611,  1991.}

\lr \khga {R. Khuri, \jnl \np, B387, 315, 1992;  \jnl \pl, B294,
325, 1992; J. Gauntlett, J. Harvey and J. Liu, \jnl \np, B409, 363, 1993.}

\lr \dukh {M. Duff, R.  Khuri,  R. Minasyan and J. Rahmfeld,
\jnl \np, B418, 195, 1994. }
\lr \susy  {R. Kallosh, A. Linde, T. Ort\'in, A. Peet and A. Van Proeyen,  \jnl
\pr, D46, 5278, 1992.}
\lr \gaun{J. Gauntlett, talk presented at the conference
``Quantum Aspects of Black Holes",
Santa Barbara, June 1993.}
\lr\gersh{D. Gershon, \jnl \pr, D49, 999, 1994.}
\lr\wil{C. Holzhey and F. Wilczek, \jnl \np, B380, 447, 1992.}
\lr\edw{E. Witten, \jnl \pl, B149, 351, 1984.}
\lr \andr{O.D. Andreev and A.A. Tseytlin, \jnl \np, B311, 205, 1988; \jnl \mpl,
A3,
1349, 1988.}
\lr \kalor { R. Kallosh and T. Ort\'in, ``Exact $SU(2)\times U(1)$ stringy
black holes", SU-ITP-94-27, hep-th/9409060. }

\lr \caa{A. Abouelsaood, C. Callan, C. Nappi and S. Yost, \jnl \np, B280, 599,
1987;
H. Dorn and H. Otto, { Z. Phys.} {C32} (1986) 599.}

\lr\ano {P. Orland, \np B428 (1994) 221. }
\lr \anoo{ M. Sato and S. Yahikozawa, \np B436 (1995) 100. }
\lr \horts{ G.T. Horowitz and A.A. Tseytlin,
\pr D50 (1994) 5204.}
\lr \hortse{ G.T.  Horowitz and A.A. Tseytlin,
\pr D51 (1995) 2896.}

\lr\sse{A. Sen and J.H. Schwarz, \pl B312 (1993) 105.
 }
\lr \rutse{J. Russo and A.A. Tseytlin,  ``Heterotic strings in a uniform
magnetic field",
hep-th/9506071.}

\lr \sixd {A. Sen, ``String string duality conjecture in six dimensions and
charged solitonic strings", hep-th/9504027; J.A. Harvey and A. Strominger,
``Heterotic string is a soliton", hep-th/9504047. }
\lr\tend{A. Dabholkar, ``Ten dimensional heterotic string is a soliton",
hep-th/9506160;
C.M. Hull, ``String-string duality in ten dimensions", hep-th/9506194.}
\lr\duf{M.J. Duff, R. Khuri and J. Lu, hep-th/9412184.}
\lr\ght{ M.J. Duff, G.W.  Gibbons and P.K. Townsend, \pl B332 (1994) 321.}
\lr \hull{C.M. Hull and P.K. Townsend, \np B438 (1995) 109.}
\lr \wite {E. Witten, \np B443 (1995) 85. }

\baselineskip8pt
\Title{\vbox
{\baselineskip 6pt{\hbox{ }}{\hbox
{Imperial/TP/94-95/54 }}{\hbox{hep-th/9509050}} {\hbox{  }}} }
{\vbox{\centerline { On singularities of    }
\vskip4pt
\centerline {spherically symmetric backgrounds   }
\vskip4pt
\centerline { in string theory }
}}

\vskip -3 true pt

\centerline{   A.A. Tseytlin\footnote{$^{\star}$}{\baselineskip8pt
e-mail address: tseytlin@ic.ac.uk}\footnote{$^{\dagger}$}{\baselineskip8pt
On leave  from Lebedev  Physics
Institute, Moscow.} }

\smallskip\smallskip
\centerline {\it  Theoretical Physics Group, Blackett Laboratory}
\smallskip

\centerline {\it  Imperial College,  London SW7 2BZ, U.K. }
\bigskip\medskip\bigskip\medskip

\centerline {\bf Abstract}
\medskip\medskip

\baselineskip8pt
\noindent

We suggest that for
singular rotationally invariant  closed string backgrounds
which need sources  for their support at the origin
(in particular, for special plane
waves and fundamental strings)
 certain `trivial'
$\a'$-corrections  (which are  usually ignored since in the absence of sources
they can be  eliminated by a field redefinition)
may play an important role
leading to the absence of   singularities  in the exact solutions.
These corrections effectively regularize the  source delta-function
 at the scale of
 $\sqrt{\a'}$. We  demonstrate that similar smearing   of the
singularity  of the leading-order point-charge  solution
 indeed takes
place in the open string theory.

\Date {September 1995}

\baselineskip 14pt plus 2pt minus 2pt

\def \ijmp {Int. J. Mod. Phys.}

\lr \fish{W. Fischler, S. Paban and M. Rozali, ``Collective coordinates in
string theory",
hep-th/9503072.}
\lr\sus{W. Fischler and L. Susskind, \pl B171 (1986) 383; B173 (1986) 262.}

\lr \tsemark {A.A. Tseytlin,
 ``Black holes and exact solutions in string theory", hep-th/9410008;
in: {\it Proceedings of the  Chalonge School on Current Topics in
Astrofundamental Physics}, September 1994, Erice,   ed. N. Sanchez (Kluwer
Academic Publishers, 1995). }

\lr \soleng {H. Soleng, ``Charged black point in General relativity coupled to
the logarithmic $U(1)$ theory", preprint CERN-TH/95-110/Revised.}

\lr\hawk{A. Lyons and S.W. Hawking, \pr D44 (1991) 3802.}
\lr \fund{C. Callan and R. Khuri, \pl B261 (1991) 363;
R. Khuri, \np B403 (1993) 335.}
\lr \dufff{ M.J. Duff, \np B442 (1995) 42; M.J. Duff and R. Khuri, \np B411
(1994) 473.}
\lr \horm {G.T. Horowitz and D. Marolf, ``Quantum probes of space-time
singularities", qg-qc/9504028. }
\lr \dah{A. Dabholkar and  J. Harvey,  \prl 63 (1989) 478.}
\lr \tsred{A.A. Tseytlin,
\jnl  \pl, B176, 92, 1986; \jnl  \np, B276, 391, 1986;
D. Gross and E. Witten, \jnl \np, B277, 1, 1986.}
\lr\john{J.H. Schwarz, ``An SL(2,Z) multiplet of type IIB superstrings",
hep-th/9508143.}


\newsec{Introduction}
Fundamental  string theory is intrinsically non-local with
an effective  space-time  cutoff of order $\sqrt{ \a'}$.
The  non-locality is reflected, e.g., in the presence
of terms of all orders  in $\a'$
 in low-energy  effective action.
It suggests  that short distance divergences should be absent not only
in the  quantum string loops but also
in the classical solutions
(both vacuum ones and the ones supported by sources).
Singularities of  leading-order solutions
may disappear  once $\a'$-corrections  are summed up.
For example, given  that the Schwarzschild solution gets non-trivial
$\a'$-modification
 at the  next to the leading order \callan, one
 may expect  the $r=0$ singularity  may be absent in the exact metric (which
may look like, e.g., as follows   \refs{\tsemark, \harms}  $\  G_{00}= -1 + {2M
\ov r}\exp (- c{ \a' M\ov r^3}), \ $ $M=GM_0$,  $G\sim L^2_{Planck}\sim \a'$).
The possibility   that higher order $\a'$-corrections may eliminate
   the   singularity  of  the spherically symmetric solutions
is supported by  analogy
with the open string theory where
the point-like singularity of the Coulomb solution  is  indeed  `smeared out'
once the Maxwell action is replaced by  the Born-Infeld plus derivative
corrections one  (see below).

One may expect that since
the  first-quantized  fundamental string is  effectively a non-local object,
the background produced by it  acting as a source  should be non-singular.
If one considers a closed string in a flat space with  a large compact
dimension $y$, the  long-distance
interactions  of the two macroscopic winding string states
($y=\s,\  t=\tau$)  can be described as a motion of one string in the
`massless' field of the  other \dah.
This corresponds to taking into account only leading-order interactions
of the winding string with  the massless string modes.
The resulting weak-field solution to  the Laplace-type  equations
can be generalized \gibb\ to the solution of  the leading-order effective
string equations  (this corresponds to taking into account  interactions
between  `soft'  massless
string modes).
This macroscopic string or  `fundamental string' (FS)  solution  \gibb\
is valid only in the weak-field region  and thus  its apparent singularity
at the   position of   the  string source ($r=0$) should not be taken too
seriously.
Indeed, near $r=0$ where the  field strengths blow up
the $\a'$-corrections
should  become important  and may eliminate the
 singularity.\foot{The fundamental string solution \gibb\ (in various
variables, see, e.g.,  \refs{\duf,\dufff})
plays  an important role in   the arguments
suggesting relations between different superstring theories in various
dimensions
\refs{\sixd,\tend,\john}. Sometimes
 an emphasis is made on singularity of the FS solution  when  represented as a
solution of the original theory  and its non-singularity when expressed in
terms
of the variables of the dual theory
(and interpreted as a soliton of the latter).
It seems that the discussion of singularity or regularity of a solution
of  the leading-order (supergravity) equations
may not have  much sense since  near the singularity
higher order corrections should be taken into account.
 What is more relevant  is whether
a given solution needs  or does not need a source for its
 support at the origin.}

One possible approach  is  to define the fundamental string background
  directly at the string  world-sheet level in terms of
 a conformal \sm  \refs{\horts,\hortse}. The resulting conformal invariance
conditions
($R_{\m\n} +...=0,$ etc.)   are satisfied at $r=0$
without  need  for a source.  It turns out that there exists a scheme in which
  the leading-order FS solution
is not modified by $\a'$-corrections and thus remains singular.
The  singularity of the FS background
 may  actually be absent  at the level of the  corresponding conformal field
theory.

Alternatively, one may    start with  the leading-order string  effective
equations
($R^{\m\n} +...=0$, etc.)  and   interpret the apparent presence of the
singularity at the origin as suggesting that the FS  solution
should be supported by a source.
Here we shall follow this   original  approach of \refs{\dah,\gibb} and
 consider
FS solution  as corresponding to the combined `field+source' action
\eqn\acti{ \hat S=S (\vp)  + I (x,\vp)  =
\int d^D x \sqrt G \ e^{-2\p} [R + 4 (\del \p)^2 - {1\ov 12} H^2_{\m\n\l} +
O(\a') ] }  $$
+ \  {1\over { 4 \pi \a' }} \int d^2 z \sqrt {\g} \ [  \del_m x^\m
\del^m x^\n G_{\m
\n}(x)  +   \ep^{mn} \del_m x^\m \del_n x^\n  B_{\m\n}(x)   +  \ \a'
R^{(2)}  \p (x)
 ]  \ . $$
The latter contains both the effective action for the background  fields  $\vp=
(G,B,\p)$ (condensates of massless string modes)  and the action
of a source string interacting with the background. Though such mixture of
actions  may look strange from the point of view of
  the  fundamental string theory,
 $\hat S$ can  be interpreted  as  representing  a non-perturbative
 `thin handle' or `wormhole'  resummation of  quantum string  theory
 {\tsmac}\ (see also \hawk).
Extrema of $\hat S$ can  be viewed as
 describing  a semiclassical approximation (both in string coupling  $e^{\p}$
and in $\a'$)
in which one `big' macroscopic string is interacting with a condensate of
 massless string modes.
The structure of the effective action $S$ is  not unambiguous: some
coefficients
in $S$ change under  local covariant field redefinitions which  preserve the
string S-matrix {\tsred}.
The important  observation  is that such redefinitions are no longer allowed in
$\hat S$ since  they transform   also the second  string source term.
The solutions corresponding to $\hat S$   will   thus  be sensitive to an {\it
off-shell }  form of the effective action $S$ which supplements $I$ in the
exponent  after a   resummation of string perturbate expansion \tsmac.
In particular, they  will  depend
on  `propagator corrections'  to  $S$ (which
are usually dropped out  since they can be eliminated  by a field
redefinition).

As we shall discuss below,  in the spherically symmetric cases where
the singularity at the origin  can be attributed to the presence of a source
one cannot  a priori  ignore  even these
`redefinable' $\a'$ corrections.
Under a certain natural assumption about the structure of these corrections
their effect will be to
  completely smooth out the singularity.
Since  (in the presence of source-related singularities)
 different off-shell extensions of $S$ may lead to different results
one  should   not thus  disregard the possibility that the singularity is
avoided in the exact solution.
It remains to be understood if there is an additional `bootstrap' condition
that
may  fix  the structure of the action/solutions in a  unique way.

 The issue of the
singularity of the FS solution will be addressed in  Section 2.
We shall     first discuss  the
spherically symmetric plane-wave solution
which is related to  the FS  background by a duality transformation.

In Section 3 we shall demonstrate that  the spherically symmetric solution
corresponding  to a point-like  source
in the open string theory is non-singular,
with the  singularity of the leading-order solution being eliminated  by
$\a'$-corrections.

Section 4 will contain some concluding remarks.

\newsec{Singular  spherically symmetric solutions in  closed string theory}

The solutions we are going to discuss below  are  special
(and simpler than, e.g., Schwarzschild) in that,
ignoring  the presence of  sources,  one  could formally  argue
 that
there exists a special scheme in which they  are not modified
by $\a'$-corrections \refs{\horts,\hortse}.
  Assuming  that one  starts from  the effective  action and
considers $G_{\m\n}$ as a  fundamental variable, i.e.
  $R^{\m\n}+ ... =0$ as the  fundamental equations,
these solutions need to be supported by sources.
As a result,
there is no longer an equivalence between
 different off-shell extensions and there is a possibility that
$\a'$-corrections can
lead to a  `regularization'  the delta-function source thus  eliminating
 the corresponding singularity.
\subsec{Plane wave solution}
We shall first discuss
the plane-wave type solution described by
the following \sm\ (${\cal R}= \four R^{(2)}$)
\eqn\fss{   L =  \del u \bd v
 + K(x)\del u \bd u +  \del x^i \bd x_i  + \a' {\cal R} \p_0 \ .   }
This model
is  formally conformal to all orders in $\a'$
provided  \guv
\eqn\lap{\Delta K=0 \ , \ \ \  \ \  \ \Delta \equiv \del^i\del_i \ . }
There are no $\a'$-corrections assuming one uses the minimal subtraction scheme
in deriving  the $\beta$-functions, or equivalently, ignores the propagator
corrections in the effective action.
The
condition  \fss\ has  the standard  plane-wave solution
$K= a_ix^i + m_{ij}x^ix^j, \ m_{ii}=0.$ It admits  (for $r >0$)
also the   rotationally  symmetric
  solution \gib\  ($K=K(r), \ r^2=x^ix_i, \ i=1,...,N=D-2$)
\eqn\hkh{ K ={ 1 + {M \ov r^{N-2}} }   \ , \ \ \  N >2 \ ,  } $$  \ \ \ K= { \
- {M \
\ln \ {r\ov r_0} } } \ , \ \ \  N=2\  ,
 $$
which
is singular at the origin
where  the components of the curvature blow up.
The conformal invariance equation \fss\ is {\it not} actually
 satisfied at the origin:
 the trace of the 2d stress-energy tensor
is  $T^m_m \sim \delta^{(N)}(x) \del u \bd u,  $
i.e.  it  receives a non-trivial contribution
which is local in space-time but
`global' on the world sheet. Thus  the
The model   is {\it not} conformal unless
some extra assumptions are made,
e.g., the line $r=0$ is `cut'  out off the space or  an external $\d$-function
source
is added.

Indeed, \lap\ is  a solution of the Poisson equation
($\m = M (N-2)$ vol$ \  S^{N-1}$)
\eqn\laps{\Delta K= - \m  \delta^{(N)}(x)  \ .  }
One may try to interpret the  $\delta$-function  term
as corresponding to a classical  string source.\foot{In \gib\ the $D=5$
solution  \hkh\  was interpreted  as a field
of an infinite string  boosted to the speed of light. However, it does not seem
to be possible to make such identification precise:
  the
configuration
$u= \s, \ v= \tau,\  x^i=0$  produces   upon the substitution into the
equations which follow  from \acti\   also a source
$\sim \delta^{(N)} (x) \int d\s d\tau \d(u)\d(v) \del_m u \del^m u $ for the
$R^{uu}$-equation  which should be satisfied trivially in the case of \fss. No
other string configuration  seems to
work either (this applies also to more general models of \hortse\ containing
$K\del u \bd u$ term with $K$ given by \hkh).
We shall ignore this complication  since it  is
absent in the case of the FS solution  we are mostly interested in. The latter
 can  indeed  be supported by a string source in a consistent manner \gibb.}
Accepting the presence of a source one is to reconsider the issue
of the $\a'$-corrections to  the l.h.s. of \laps.
In general, given a background \fss\ admitting a covariantly constant null
Killing vector one still has a freedom of the  following field redefinitions
(note that all  non-trivial   second-rank  tensors  involving contractions of
curvature tensors vanish on this background)
\eqn\rede{G_{\m\n} \to G_{\m\n}
+ b_1\a' R_{\m\n} + b_2 \a'^2 D^2 R_{\m\n} + ...+ b_n \a'^{n+1 } D^{2n }
R_{\m\n}
 + ... \ . }
This implies the replacement of the $ R_{\m\n }=0 $ equation \lap\ by
\eqn\yyy{ R_{\m\n}
  + d_1 \a' D^2R_{\m\n} + ... + d_n \a'^n D^{2n} R_{\m\n}
 + ...  \equiv
f(\a' D^2) R_{\m\n} =0 \ , }
which has the  non-vanishing  $uu$-component being equivalent to
\eqn\bee{ f(\a'\Delta) \Delta K=0 \ ,  \ \ \ \ \ \ f(z) = 1 + d_1 z + d_2 z^2 +
... \ .
 }
Eq. \bee\ is the analog of \lap\ in a generic scheme.
The same function $f$   appears also in  the kinetic term in $S$.
 The simplest  and most natural `non-minimal'
scheme choice  corresponds to  $f(z)  = e^{-cz}, \ c=-d_1$, i.e.,
\eqn\bett{ e^{-c\a'\Delta} \Delta K=0 \ . }
Indeed,  to find
 the condition of conformal
invariance of \fss\  one
 first integrates over $u,v,$
obtaining the  effective scalar (`tachyonic') vertex  $K(x)\del u_0 \bd u_0$
 ($u_0$ is the  background    value of $u$).
The condition of the zero anomalous dimension of this interaction term
then leads to \lap.  This is,
however,  a sufficient but not  necessary condition  for
 conformal invariance. The  divergent term in the 2d
effective action is
  $ \exp (-\a' \log \ep\ \Delta)K(x)\del u_0 \bd u_0 $
 ($\ep$ is  a 2d  UV cutoff).
Taking the derivative over   $\log \ep$  and setting $\log \ep = c$ ($c\not=0$
corresponds to a   non-minimal subtraction  scheme in which   higher-order
tadpoles contribute to the $\beta$-function)
leads to   \bett.

One  usually expects that physical properties of  perturbative
vacuum string solutions
should not depend on a choice of
parameters which represent the freedom of off-shell continuation.
Indeed, the perturbative solutions of \bee\ or \bett\
are the same  as of \lap.
This, however, is no longer true in the presence of {\it sources}.
In particular, these equations  are {\it not} equivalent
in the case of the singular  leading-order
solution \hkh\  which
needs a $\d$-function source for its support.
In what follows we shall use the simple
 exponential choice  for  the function $f(z)$ \bett,
\eqn\beep{   e^{-c\a'\Delta} \Delta K= -\m  \delta^{(N)}(x)  \ .  }
Rewriting \beep\ as
\eqn\bep{    \Delta K= -\m  \delta^{(N)}_{\a'} (x)  \ , } $$
\delta^{(N)}_{\a'} (x) \equiv e^{c\a'\Delta}\delta^{(N)}(x)  $$ $$
=\  \int {d^N k \ov (2\pi)^N }\  \exp ( ikx - c\a' k^2)
= (4\pi c\a')^{-N/2} \exp (- {x^2 \ov 4c\a'})  ,   $$
one can  also interpret the replacement of \laps\ by \bep\ as
as a result of
 `smearing'  of the $\d$-function source at the string  scale
 $\sqrt{ \a'}$.\foot{This smearing can be attributed to the
quantum fluctuations of the first-quantized string source
which make the interaction of the string with the background fields non-local.
Indeed,  taking the expectation value of the $\d$-function
$\d (x(\s,\t)-x)$  corresponds to replacing  $e^{ikx} $ by
  $e^{ikx - c\a' k^2} $, \ $c= \log \ep$ in its Fourier representation.}
This is analogous  to  the `regularization' (at the scale of  the inverse Higgs
mass) of a   similar  $\d$-function
 source in the case of  the Abrikosov-Nielsen-Olesen
string (see, in particular, \refs{\ano,\anoo}).
There
the quantum effects  transform  the   $\d$-function which describes
the coupling of  the `axion'  $B_{\m\n}$  (which is dual to angular part of the
Higgs
field)
 to the string into  a non-singular function,  e.g, $ (2\pi)^{-4} \int d^4 k
\exp (ikx -  c \Lambda^{-2}k^2 )$ \anoo.
Note also  that  the transformation  \rede\ or $K \to f(\a' D^2) K $
which relates  \lap \  and  \beep\  can be interpreted  as a
modification of the field--string coupling in the
 string  source action  term in \acti.

Assuming $c > 0$  the solution of \bep\ is  {\it regular}
 at $r=0$ and  reduces to \hkh\ at  large $r$.
For example, in the case of $D=5$, i.e. $N=3$ it  is given by
(cf. \hkh)
\eqn\ppp{ K= 1 + {M\ov r}  {\rm erf }  \big({r\ov 2\sqrt{c\a'}}\big) \ , \ \  \
\  \ \ \   {\rm erf} (b) = {2\ov \sqrt \pi}\int^b_0 dz \ e^{ -z^2}\ . }
The  conclusion about regularity  then applies also  to the $a=\sqrt 3$
`Kaluza-Klein'
extreme electric black hole solution which  is obtained from  the  $D=5$
solution
\fss\ by dimensional reduction \gib\ (${\cal A}_\m$ is the vector field and
$\s$ is the modulus)
\eqn\exttt{ ds^2_4= - K\inv (r) dt^2 +  dr^2 + r^2 d\Omega^2_2 \ , \ \
\  {\cal A}_t = K\inv (r)\ , \ \  \ \s =  {\ha}\  \ln\ K (r) \ . }
To summarize, starting with a  generic  choice of the effective action
and  studying how the leading-order rotationally-symmetric solution
\hkh\ is modified by $\a'$-corrections one learns
that because of the necessity to introduce a source at the origin
the $\a'$-corrections cannot be ignored
and may completely eliminate
the singularity.

\subsec{Fundamental string solution}

The FS solution \gibb\ is described  by the following 2d \sm action  \horts
\eqn\fsss{   L = F(x) \del u \bd v
+  \del x^i \bd x_i +    \a' {\cal R}
\p (x) \ ,   \ \ \ \ \p = \p_0  + \ha \ln  F (x)\ . }
This action is related to \fss\ by the duality transformation
(if
we set $u=y-t, \ v=y+t$ then the duality $y \to \tilde y$ leads to \fss\ with
$u= \tilde y$, $\ v=t$, \  $K=F\inv$).
Expanding near a  2d background
$(u_0,v_0,x_0)$ one finds that the condition of conformal invariance of \fsss\
is
\eqn\cond{ \Delta  F  + 2 F\inv \del^i F \del_i F =
F^2 \Delta F\inv =0 \ ,  }
which is equivalent to $R_{\m\n } + ...=0$.
As was argued in \horts,\ there exists a scheme in which
 \cond\ represents the  exact
conformal invariance condition to all orders in $\a'$.

If, instead, one  introduces 2d
sources $J_u,J_v$ for $u,v$ and then integrates $u,v$  out,  one ends up with
the  effective `tachyonic' vertex $ F\inv (x)  J_u \bar J_v $
so that the condition of conformal invariance is just
\eqn\coo{ \Delta F\inv =0 \ ,   }
as  in the  dual model \fss\  with $K=F\inv$.
 The condition \coo\ follows   from $R^{\m\n} + ...=0$
(which is the same equation that is obtained from the effective action by
varying $G_{\m\n}$). It
is  equivalent  to \cond\
provided $F$ is non-vanishing everywhere.
This is no longer true in the
 case of the singular  FS solution  \gibb\
with  $F\inv =K$  given by  \hkh. Since  $F$ does vanish (for $D>4$) at $r=0$
eqs. \cond\ and \coo\ are not equivalent: while \cond\ holds  everywhere,
\coo\ needs to be supported by a source at the origin.

Following  \gibb\  let us assume  that the source  representing a
macroscopic  string state should
indeed be  added to the  r.h.s. of \coo.
The FS background is then a  consistent leading-order solution
of the equations corresponding to the action \acti\
supported by the source produced by  the string configuration $u= \s + \t, \
v= \s-\t, \ x^i=0$ (which remains a solution of the string equation also
in the case of the non-trivial background \fsss).
As in the plane-wave case discussed in the previous subsection,  we are then to
address  the issue of $\a'$-corrections to the  leading terms in the effective
action, i.e. to the l.h.s. of eq. \coo.
As discussed above,
 the  intrinsic non-locality of the 1-st quantized string
suggests that in the presence of a source one should replace the
$\Delta$-operator in \coo\  by a `regularised' expression
$f(\a'\Delta) \Delta$ as in \bee.
 For the  simplest  exponential  choice of $f$
the resulting  equation is then  equivalent to \beep\ and
thus the $r=0$ singularity of the FS solution is smeared out.
Let us stress  again that  in the presence of a source
it is no longer true that   different choices  of the function $f$ or even of
the constant
$c$ in \bett\ lead to physically equivalent solutions.

Similar   conclusion  should apply  also  to the  generalisations
\refs{\garf,\sen,\wald,\ghrw}
 of the  FS model \fsss,
in particular, to  the following one ($n= \pm 1$) \hortse\foot{Let us note also
that all  charged fundamental  string solutions
discussed in \sen\ and the  first reference in \sixd\
are described by the  special  chiral  null model  \hortse\
($y^I$ are internal toroidal coordinates)
$\ L= F(x) \del u [\bd v + K(x) \bd u + 2 A^I_u (x) \bd y^I] +  L_{IJ} \del y^I
\bd y^J + \del x^i \bd x^i  + \a'{\cal R}\p (x) , $
or its straightforward  `heterotic' generalisation in which part of $y^I$
are represented  by chiral scalars (see \refs{\sse,\rutse}).}
 \eqn\fsss{   L = F(x) \del u \bd v
+  n \del x^i \bd x_i + \a'{\cal R}\p (x) \ ,   \ \ \ \ \p = \p_0  + \ha \ln  F
(x)\ .  }
Such $D=5$, $n=1$ model describes (upon dimensional reduction to $D=4$)
the $a=1$ extreme electric black hole background
\eqn\did{ ds^2_4
 = - F^2 (r) dt^2 + dx_idx^i \ ,  \ \ \ \
\cA_t=-\cB_t = F(r)\  , \ \ \ \ \e{2\p}   = F (r)\ .   }
It is thus plausible that when the   $\a'$-corrections  are
included, this background
is  no longer   singular at $r=0$.

\newsec{Regularity of the  point-charge  solution in  the  open string theory}
In the examples of solutions discussed above the
$\a'$-corrections which were responsible for smearing of
 the singularity where of `propagator' or `redefinable' nature.
In general, there are also non-trivial $\a'$-corrections that cannot be
eliminated
by a field redefinition even in the  absence of sources.
The example of the open string theory discussed below suggests that
 their effect is also to  smooth out  the
singularity  of  a  leading-order solution.

Namely, we  shall show that
 the point-charge singularity of the Maxwell theory is  absent
 in the  open string theory.
The tree level (disc) term in the abelian vector field  effective action of the
 open (super)string theory  has the following structure
\refs{\frtse,\caa,\andr}
\eqn\bor{ S= e^{-\p_0} \int d^Dx \big[\sqrt {-\det( \eta_{\m\n } + 2\pi \a'
{\cal F}_{\m\n})
} +  \a'^3 f_1(\a'{\cal F})  \del {\cal F} \del {\cal F} + ... \big] \ ,  }
where $f_1 = a_1 \a'^2{\cal F}^2 + ... $.
Let us first ignore all field strength derivative  $O(\del {\cal F})$ terms
and consider just the first Born-Infeld term.
Adding a  charged open string source  to \bor\
(which can be represented by a point-particle source term $\sim Q A_0 (x)
\delta^{(D-1)} (x) $ since the open string is charged only at its ends)
one may  find  the
 corresponding electric field.

 A remarkable property of the Born-Infeld Lagrangian is that while in the
Maxwell
theory the field of a point-like charge is singular at the origin and  has
 infinite energy, in the Born-Infeld case   the field is regular at $r=0$
(where the electric field takes its maximal  value)
and its total energy is  finite \born.
For example,  in  $D=4$, \
 $ \sqrt {-\det( \eta_{\m\n } + 2\pi \a' {\cal F}_{\m\n})}
= \sqrt {1- (2\pi \a' E)^2} $ so that the  analogue
of the Maxwell equation is
$\del_r (r^2  { \cal D }) \sim Q\delta (r) , \ \ \
 {\cal D} \equiv   E/\sqrt {1- (2\pi \a'E)^2}$.
The solution is  $ {\cal D} = Q/r^2 $, i.e.,
\eqn\fii{  \  E\equiv {\cal F}_{rt}={  Q\ov \sqrt{r^4 + r_0^4} } \ , \ \ \ \ \
\  r_0^2\equiv
2\pi\a'  Q \ . }
{}From the point of view of the
distribution of the electric field ($\rho=$ div $  E / 4\pi$) the source is no
longer point-like  but  has an effective  radius  $r_0 \sim \sqrt {\a' Q}$.

The electric field is approximately constant  ($E \sim Q/r^2_0 \sim 1/\a' $) in
the region $ 0\leq r<r_0$.
Its  derivative vanishes at $r=0$ and is suppressed by a power of $Q$
near  $r \sim r_0$ ($\ \del_r E \sim Q/r^3_0 \sim {\a'}^{-3/2} Q\inv$).
That means that the effect of the derivative terms in \bor\ should be small,
i.e.
the qualitative conclusion about the regularity of the static spherically
symmetric point  source solution applies  to the  {\it full effective action}
of the open string theory.\foot{Moreover,  if $Q$ is large  the derivative
corrections to the effective action  do not significantly modify the   form of
the Born-Infeld solution \fii.}

\newsec{Concluding remarks}
We  have argued that
in the cases when  there are   no `non-trivial' $\a'$-corrections but there
is a source at the origin  (so that the string equation reduces, e.g.,  to the
Poisson  equation with a $\delta$-function source)
one  may not be able to ignore  the effect of the  `trivial' $\a'$-corrections.
Then
the  exact solution   may turn out to be non-singular.
 The `smoothening'
of the  singularity of the solution
corresponding to a fundamental string source  seems  to be
 natural: the quantum string has a built-in cutoff
at the space-time scale $\sqrt {\a'} $ and like the Abrikosov-Nielesen-Olesen
 string  (for a finite Higgs mass)
 should be
non-singular.

The  suggested absence of singularities in the  solutions produced by local
string sources  is supported  by the example of the open string theory
where  the $\a'$-corrections eliminate the singularity  of  the  Coulomb
solution.
This strengthens  the expectation   that  the exact version of the
Schwarzschild solution should also be non-singular.

There are many open questions remaining.
One is  the existence of a well-defined CFT  description behind such solutions
with sources.
Another is about
 a formulation of conformal invariance conditions in the presence of  sources
which  effectively chooses the `right' form of the off-shell effective action
(or, as in the absence of  sources,
 is not sensitive  to
 field redefinitions).
Given that \acti\ corresponds to a resummation of the standard string
perturbation theory \tsmac\
such formulation should be related to the
 attempts to extend the condition of conformal invariance  to the string loop
level
\refs{\sus, \fish}.
It should be useful also to  understand  the reason behind  the
strange  feature of the
 fundamental string  background  that  it needs or does not need  a source  for
its support depending
on the form  of the Einstein  equations (with raised or  lowered indices)
  one  chooses to  solve
 and depending on the metric one  uses (string  metric or
 the one rescaled by a function of the dilaton).\foot{The change of variables
(conformal rescaling of the metric)
used, e.g., in \refs{\duf,\ght}
 seems  to   eliminate or introduce a source  in the leading-order
string equations. The  Einstein
equation  with  raised indices  and a source in the r.h.s.
$R^{\m\n} + ... \sim  \int d \s  d \tau  \delta (x - x (\s,\tau)) \del x^\m \bd
x^\n$
is  not invariant under  the  conformal rescalings of the metric.
 Rescaling the metric  by a  factor which  vanishes at the origin  one  may
effectively  drop out   the source contribution.}

\bigskip\bigskip
{\bf Acknowledgements}

\noindent
I am grateful to  J. Russo, K. Sfetsos and K. Stelle
for useful discussions.
I acknowledge the hospitality of  the CERN  Theory Division
while this work was in progress  and the   support
of PPARC, ECC grant SC1$^*$-CT92-0789
and NATO grant CRG 940870.

\vfill\eject

\listrefs

\end